\documentclass[aps,twocolumn,prb,showpacs]{revtex4}
\usepackage{graphicx}
\usepackage{epsfig}
\usepackage{amsmath}
\usepackage{graphics}
\begin{document}

\title{Nature of  well-defined conductance  of amine anchored molecular junctions}

\author{Zhenyu Li}
\author{D. S. Kosov}

\affiliation{Department of Chemistry and Biochemistry, University of Maryland, College Park, MD  20742}

\begin{abstract}
Amine terminated molecules show well behaved conductance 
in  the scanning tunneling microscope break-junction experimental measurements.
We performed density functional theory based electron transport calculations to explain the nature of this phenomenon.  
We find that amines can be adsorbed only on  apex Au atom,  while thiolate group can be attached 
equally well to undercoordinated and clean Au  surfaces. Our calculations show that only one adsorption geometry is sterically and energetically possible
for amine anchored junction  whereas three different adsorption geometries with very distinct transport properties  are almost equally probable for thiolate anchored junction. We   calculated  the conductance as a function of the junction stretching when the molecules are pulled by the 
scanning tunneling microscope tip from the Au electrode. Our calculations show that  the stretching of the thiolate anchored junction during its formation is accompanied by  significant electrode geometry distortion. 
The amine anchored junctions exhibit very different behavior -- the electrode  remains intact when the scanning tunneling microscope tip stretches the junction.
\end{abstract}

\maketitle

\section{Introduction}

Considerable experimental and computational efforts have been devoted in recent years to molecular 
electronics.\cite{ratner-nitzan} 
%Although these efforts have been spurred on by the potential for breakthrough technological applications, they have already revealed  new fascinating scientific problems in interface chemistry. 
The main goal in molecular electronics is the formation and  characterization of
nano-electronic device in which a molecule plays the role of  a conducting element. Molecular junctions are formed by placing molecules in thin, sub-nanometer gap between metal electrodes.
It is  assumed that under certain conditions a single molecule bridges the gap and conducts the electricity. Although there has been a significant progress towards experimental determination of a single molecular conductance, there is growing awareness that the number of molecules which bridge the gap is not a well characterized experimental quantity.  Indeed, it is difficult to prove experimentally without 
uncertainty that  the measured value is the single molecular conductance rather than the conductance through  molecular array.

In the experimental technique, which was pioneered by Tao {\it at al.}\cite{tao-science,tao-jacs, Xiao0467}
and then used by Venkataraman \emph{et al.}\cite{Venkataraman0600,Venkataraman06-nat} for amine terminated molecules, the
 junctions are formed by repeatedly crashing an Au scanning tunneling microscope (STM) tip into and moving it out of Au surface in a molecular solution.
 The recent experiment  of Venkataraman \emph{et al.} \cite{Venkataraman0600} demonstrated that amine anchored molecular  junctions exhibit well-defined and experimentally stable transport properties, namely 1,4-benzenediamine (BDA) 
has 2-3 orders smaller variations of the conductance between different experimental measurements 
than  1,4-benzenedithiolate (BDT).\cite{Venkataraman0600} This is a counterintuitive observation. The thiolate-Au chemical bond is considerably stronger than the amine-Au bond. Therefore, one may expect that the stronger  molecule-electrode link would reduce the stochastic switching and result into more stable 
molecular junction. Moreover, the weak amine coupling to the electrode implies the decreased junction "transparency" for electron current.
What is the reason for well-defined, stable conductance of amine anchored single-molecular junctions? Why do amine terminated molecules tend to form single molecular junctions?
These are the questions which interest us here.

Our paper is organized as follows. In section II, we present computational and theoretical methods. In section III, we address three important aspects of the interface chemistry and their roles in the electron transport: (i) possible adsorption sites  for amines and thiolates  on clean and undercoordinated Au(111) surfaces, (ii) sensitivity of electron transport to the interface geometry, 
(iii) conductance evolution as the junction is formed and stretched by the STM tip. 
In section III, we also analyze and discuss these results and elucidate the nature of the well-defined conductance of amine terminated molecular junctions. Section IV summarizes the main results of our study.

\section{Computational Methods}

We used three computer programs in our calculations: SIESTA code \cite{Soler0245} to study adsorption of molecules on Au surface, 
ATK package \cite{Brandbyge0201, ATK} to compute transmission spectra for all possible adsorption geometries, and 
our own code for "on-the-fly" conductance calculations to model STM break junction experiments. In all our calculations we use the same level of electronic structure theory. Namely, we apply density functional theory (DFT)  with general gradient approximation (GGA-PBE) for 
the exchange-correlation potential.\cite{Perdew9665} 
We use  numerical single-$\zeta$ with polarization (SZP) for Au and the rest of the atoms are described 
by numerical double-$\zeta$ with polarization (DZP) basis sets.\cite{Soler0245} The core electrons are modeled with 
the Troullier-Martins nonlocal pseudopotentials.\cite{Troullier9193}

\subsection{Conductance calculations for static junctions}
ATK program is based on the combination of the non-equilibrium Green's functions (NEGF) and 
DFT.\cite{Brandbyge0201, ATK}  We describe below the details of the ATK program relevant to our applications.
The program uses the surface DFT calculations to obtain the electrode self-energy.\cite{Brandbyge0201} 
The matrix product of the Green's function and the imaginary part of the left/right electrode self-energy is used to define  the spectral density. The non-equilibrium, 
voltage-dependent density matrix  is computed from the spectral density. Then the density matrix is converted into non-equilibrium electron density.
The non-equilibrium electron density yields   the Green's function for the scattering region. The Hartree potential is determined by the solution of the Poisson equation with voltage-dependent boundary conditions. 
This iterative step [Green's function $\rightarrow$ nonequilibrium electron density $\rightarrow$ Green's function] is repeated until the self-consistency is achieved.
Then, the  transmission spectrum is calculated 
by the standard equation.\cite{Brandbyge0201}  

\subsection{Junction formation and stretching: on-the-fly conductance calculations}
We would like to model the experiments where
the  molecular junctions are formed by repeatedly crashing an Au STM tip into and pulling it out of Au surface in a molecular solution.\cite{tao-science,tao-jacs,Xiao0467,Venkataraman0600,Venkataraman06-nat}
To model these experimental conditions  we developed the computational method which calculates molecular conductance "on-the-fly". Namely, 
we performed conductance calculations along the trajectory, which is 
computationally obtained by pulling  the molecule from the Au electrode.  

The  geometry evolutions of the molecular junctions
are obtained by  constrained geometry relaxation method.\cite{Kruger0351}   
The molecular junction is aligned along the z-axis. The  $z$ coordinate of the top C atom is incremented by 0.05 \AA $ $
 on every step and this stretch of the junction is followed by the   constrained geometry relaxation.
 In the constrained geometry relaxation, the $z$ coordinate of the central bottom Au atom of the electrode cluster and the top C atom of the phenylaime or phenylthiolate molecule are fixed and the rest of the system is fully optimized.  The upper electrode is the mirror image of the bottom one.
 The junction evolution  is represented by the trajectory composed of the optimized geometries. It becomes
  the true molecular dynamics trajectory in the limit of extremely slow pulling speed. 
 Previous studies found  
that the constrained geometry relaxation method gives the results consistent with Car-Parrinello
molecular dynamics simulation of pulling molecules from surface. \cite{Kruger0202, Kruger0351}

For every step along the trajectory the conductance is computed by the method, which we implemented in the SIESTA program. \cite{Soler0245}  We compute the  self energy by the level broadening model, which has been successfully applied to molecular junctions and metal nanowires. \cite{Tada0450, Li0647} In this model, electrode energy levels are broadened  to obtain the density of states, which is related to the imaginary part of self energy.  Our transport implementation  avoids the computationally expensive Cauchy principal value numerical integration and  uses the Lorentz broadening instead, as was suggested in our non-orthogonal Wannier-type atomic orbital  electron transport calculations.\cite{Li0647}   
The Lorentzian broadening $\sigma$ 
in the electrode density of state is equivalent to positive infinitesimal $\sigma$
in the electrode Green's function:\cite{Li0647}
\begin{equation}
\mathbf{g}= [(E+i\sigma)\mathbf{S}-\mathbf{H}]^{-1}\;,
\end{equation}
where $\mathbf{S}$ is the overlap matrix and $\mathbf{H}$ is the electrode Hamiltonian.
The conductance of monoatomic gold chain attached to two gold clusters is 
$G_0= 2e^2/h$,\cite{agrait03} and the 
broadening parameter $\sigma$ is fitted to reproduce this value. 
This fitting leads to  $\sigma$=0.5 eV and we use this value in all our calculations. This computational method enables us to perform conductance calculations for hundreds of sample geometries from a junction evolution trajectory.

\section{Results and Discussion}
\subsection{Phenylamine and phenylthiolate adsorption on clean and undercoordinated gold surfaces}

First principles studies of adsorption of different amine or thiolate terminated  molecules on Au(111) surfaces have been reported.\cite{Bilic0281, Lambropoulos0277, Zhou0605, Bilic0508} 
It was predicted that ammonia is adsorbed on  the atop site.\cite{Bilic0281}  The cluster model calculations show that the adsorption energy  depends significantly on the employed theoretical method. \cite{Lambropoulos0277}  Phenylthiolate is attached to the bridge site and the phenyl ring is strongly tilted with respect to the Au surface.\cite{Bilic0508} Although previous studies are very informative, a comparative study of adsorption of amines and thiolates on Au surfaces in the context of electron transport calculations is necessary. 

We represent Au(111) surface by unit cell, which is  periodically repeated in two directions. 
The unit cell consists of five (4 $\times$  4) layers of Au atoms. 
The undercoordinated gold surface is modeled by adding an extra apex atom above the Au(111) hollow site.

\begin{figure}
\includegraphics[keepaspectratio,totalheight=12cm]{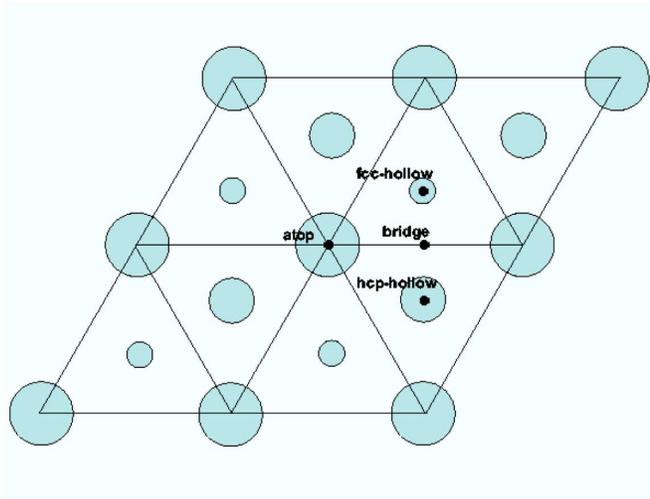}
\caption{ 
Diagram which shows possible adsorption sites on  Au(111) surface.  The Au atoms are represented by circles of different sizes. 
The smaller the spheres,  the farther from the layer the Au atoms are.
}
 \label{fig:sites}
 \end{figure}
  \begin{figure}
  \begin{center}
\includegraphics[keepaspectratio,totalheight=18cm]{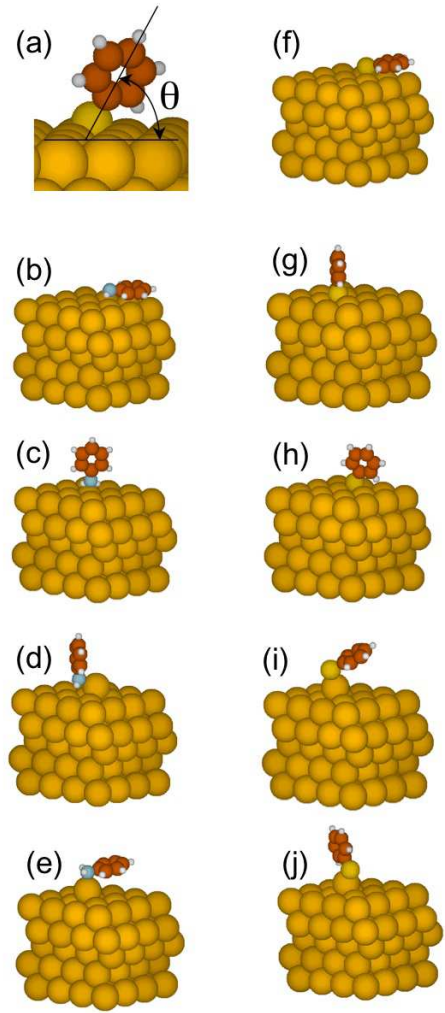}
\caption{ 
adsorption geometries for phenylamine (b-e) and phenylthiolate (f-j) on the clean (b,c,f,g,h) and undercoordinated (d,e,i,j)  Au(111) surfaces. (a) defines the angle $\theta$, which is used in Table I.
}
 \label{fig:geo}
 \end{center}
\end{figure}
\begin{table}
\caption{ Adsorption geometries of phenylamine and phenylthiolate. $N_{\rm{Au}}$ is the number of the nearest neighbor Au atoms for  the N atom  or the S atom.  
$d_{\rm{Au}-\rm{N}}$ and $d_{\rm{Au}-\rm{S}}$ are the mean distances for the $N_{\rm{Au}}$ nearest neighbors of the N atom and the S atom respectively.
$\theta$ is the angle between the phenyl ring and the surface (Fig.\ref{fig:sites} a). } \label{tbl:adsorb}
\begin{tabular}{ccccc}
\hline\hline
\multicolumn{5}{c}{phenylamine} \\
adsorption geometry   & $N_{\rm{Au}}$ & $d_{\rm{Au}-\rm{N}}$(\AA) & $\theta$(degree) & AE (eV)   \\
\hline
Fig.\ref{fig:geo} (b) & 1 & 2.90 & 0.83 & 1.92  \\
Fig.\ref{fig:geo} (c) & 2 & 3.05 & 87.5 & 0.68  \\
Fig.\ref{fig:geo} (d) & 1 & 2.34 & 87.9 & 1.18  \\
Fig.\ref{fig:geo} (e) & 1 & 2.30 & 12.0 & 0.92  \\
\hline\hline
\multicolumn{5}{c}{phenylthiolate} \\
adsorption geometry & $N_{\rm{Au}}$ & $d_{\rm{Au}-\rm{S}}$(\AA) & $\theta$(degree) & AE (eV) \\
\hline
Fig.\ref{fig:geo} (f) & 2 & 2.56 & 8.12 & 3.64  \\
Fig.\ref{fig:geo} (g) & 3 & 2.44 & 89.2 & 2.57  \\
Fig.\ref{fig:geo} (h) & 3 & 2.51 & 63.7 & 2.72 \\
Fig.\ref{fig:geo} (i) & 1 & 2.31 & 31.5 & 2.72  \\
Fig.\ref{fig:geo} (j) & 1 & 2.31 & 51.3 & 2.75  \\
\hline\hline
\end{tabular}
\end{table}

We perform geometry optimizations for phenylamine and phenylthiolate molecules adsorbed on 
clean and undercoordinated  Au(111) surfaces. A final optimized  structure is sensitive to  initial guess for  geometry optimization. We choose various starting geometries  to explore all possible adsorption sites. Fig.\ref{fig:sites} shows atop, bridge and hollow sites on Au(111) surface. There are two different hollow sites on Au(111) surface: fcc-hollow  and  hcp-hollow sites. We initially align the phenyl ring of the phenylamine and phenylthiolate along the Au-Au or the Au-hollow directions. The phenyl ring of phenylamine and phenylthiolate  can be perpendicular or tilted with respect to the Au surface. For tilted phenylamine, the initial orientation is chosen in such a way that the N lone pair points perpendicular to the surface. 
We choose angle 30$^{\circ}$ for  the initial  phenylthiolate tilted orientation.
The coordinates of the Au atoms in  the two bottom layers are constrained to theoretical bulk geometry with lattice vector 4.132 \AA\ . This  lattice vector is obtained by performing a series of bulk calculations with different lattice parameters and fitting the energies to the third-order Birch-Murnaghan equation of state. \cite{Birch4709}
We performed  two subsequent geometry optimizations to get equilibrium structures for the adsorbed molecules.
 In the first geometry optimization, the N atom or the S atom (and  the Au apex atom when  the surface is undercoordinated) are fixed on a specific adsorption site and can only move in the $z$ direction, which is perpendicular to the surface. In the second one, all atoms except the two bottom Au layers are fully relaxed.
 
Having  obtained the optimized geometries for the molecules adsorbed on the surface, we compute adsorption energy (AE) by subtracting the energy of the 
total system from the sum of energies of  the surface and the molecule.\cite{ae-bdt}
Fig.\ref{fig:geo} and Table I summarize the results of the geometry optimizations. 
Due to the failure of DFT GGA exchange-correlation functionals to account for van der Waals interactions, the actual AEs may be somewhat higher  than the calculated values.
AE is mainly determined by the orientation of the phenyl ring with respect to the surface. When the phenylamine is adsorbed on the clean Au (111) surface its phenyl ring either lies almost flat on the surface 
(geometry (b) on Fig. \ref{fig:geo}) or stays vertical to the surface (geometry (c) on Fig. \ref{fig:geo}).
 AE of geometry (b) is about 1.2 eV larger than that of geometry (c). This can be expected since there is strong attractive  interaction between the molecular $\pi$-electrons and  the metal surface image charges, which favors parallel to the surface phenyl ring orientation.  The phenyl ring lies  above the three-Au triangle on the Au(111) surface.  The differences in the adsorption positions
 of the N atom cause  only small AE variations of about 0.1 eV, if the phenylamine is attached vertically to the surface. When the
phenylamine is adsorbed on undercoordinated Au(111) surface, the phenylamine adopts vertical (geometry (d) on Fig. \ref{fig:geo}) or tilted  (geometry (e) on Fig. \ref{fig:geo}) configurations. The N  lone-pair points to the apex Au atom in both cases. AE of the vertical phenylamine (1.18 eV) is 0.26 eV larger than that of the tilted one, and it is 0.50 eV larger than AE of the most stable non-lying adsorption geometry on  clean Au(111) surface.

The geometry optimizations of  phenylthiolate on  clean Au(111) surface also yield adsorption structures where the molecule either lies almost flat on 
the surface (f), or it is vertical (g), or it is tilted  (h) with respect to the surface. The lying phenylthiolate is not perfectly parallel to the surface, because the attraction between the S atom and the surface pulls the  atom closer to the surface. In agreement with previous calculations,\cite{Bilic0508} the S atom  appears near the bridge site in the lying geometry.  If phenylthiolate is adsorbed on the apex Au atom, the molecule is always tilted with respect to the surface.
For phenylthiolate, there are two possible  tilted geometries. The phenyl ring plane is tilted in the first geometry ((i) on Fig. \ref{fig:geo}), and the second geometry has the vertical  phenyl ring but the tilted S-C-C axis ((j) on Fig. \ref{fig:geo}). The AE difference of these two possible tilted geometries is within 0.1 eV. The lying  adsorbed molecule also gives the largest AE for phenylthiolate.  But unlike phenylamine,  the tilted adsorption geometries of phenylthiolate on clean and undercoordinated Au(111) surface have very similar AEs, which are about 0.9 eV smaller than that for the lying geometry and 0.15 eV larger than AE for the vertical geometry.
%{ \it Absolute AE value for phenylthiolate depends on what happens with the released H goes after adsorption. If we suppose H forms H$_2$, which gives the upper limit of AE, the most stable non-lying adsorption geometry of phenylthiolate gives an AE 1.57 eV larger that of phenylamine. why don't you compute just the energy of the Hydrogen atom instead H2??}

\subsection{Electron transmission dependence on adsorption site geometry}

\begin{figure}
\includegraphics[keepaspectratio,totalheight=7.5cm]{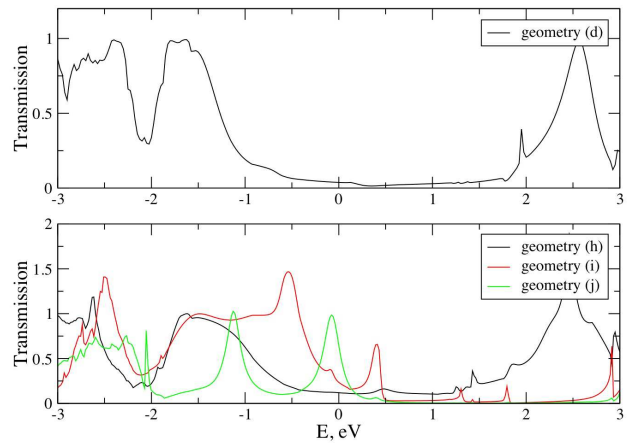}
\caption{ 
Transmission spectra of  BDA (upper panel) and BDT (lower panel) computed for different adsorption geometries. Fermi energy of the electrodes is shifted to zero. The junction geometries (d), (h),(i) and (j) refer to Table I and Fig.\ref{fig:geo}.
} \label{fig:trans}
\end{figure}
 \begin{figure*}
\includegraphics[keepaspectratio,totalheight=14cm]{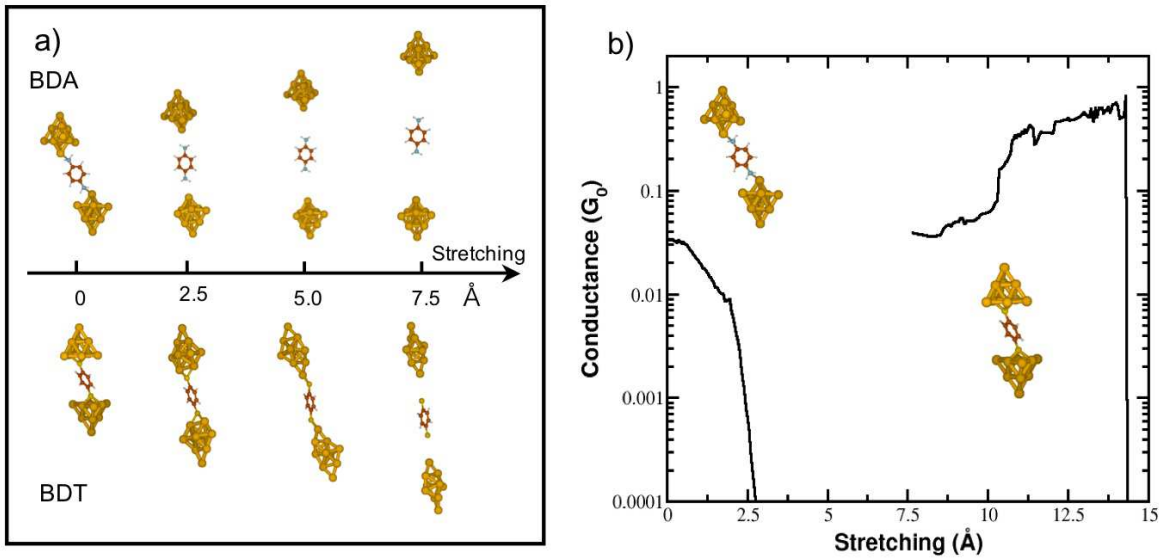}
\caption{a) Snapshots of the junction geometries.
b)
Conductance traces of the BDA (left curve) and the BDT (right curve) junctions. The starting point of conductance traces for the BDT junctions is shifted along the  $x$ axis to  7.5 \AA\ . Inset: the starting junction geometries. 
} \label{fig:stretch}
\end{figure*}

 Fig.\ref{fig:geo} and Table I list the most probable  geometries for the adsorption of phenylamine and phenylthiolate on Au(111) surface. These configurations are obtained from the geometry optimizations described in the previous section. The lying  on the surface molecular junctions (Fig.\ref{fig:geo}  (b), (e), (f)) are excluded from the transport calculations, since  it is difficult to form them experimentally. Our calculations show that their conductance is close to or  even larger than the conductance of a gold monoatomic wire, $G_0= 2e^2/h$. It means that the lying on the surface junctions  are also undetectable in the STM break-junction experiments.
Three different geometrical arrangements (h), (i), (j) are energetically and sterically possible for BDT junction to the Au surface, whereas only one geometry  (d) is available for the BDA junction. 
Given these optimized adsorption  geometries, the molecular junctions are constructed by mirror addition of the other electrode and the corresponding anchoring group. The conductances of these four junctions are  calculated by NEGF/DFT   as is implemented in the ATK package.\cite{Brandbyge0201, ATK} Transmission spectra of the BDA and the BDT junctions  are plotted in Fig. \ref{fig:trans}. 
The BDA transmission is featureless in the vicinity of the Fermi energy.
 Fig.\ref{fig:trans} shows very distinct  transmission spectra for three geometrically different BDT junctions  which have similar AEs.
 The BDT junction with geometry (h) does not have any transmission resonances near the electrode Fermi energy.  The junction with geometry (i) shows two resonances: one broad resonance is below and one resonance is above the Fermi energy. The 
 most "transparent" for electrons junction is the junction with geometry  (j). It has one peak  located at the Fermi energy.
 The transmission spectra for the BDT junctions are in quantitative agreement with the results of density functional theory based electron transport calculations 
 reported by other authors.\cite{xue-ratner,evers04,ke-baranger-yang}
 
The Landauer formula\cite{Datta95}
\begin{equation}
G=G_0 T(E_F)
\end{equation}
relates the conductance $G$ with  the  transmission function $T(E_F)$ computed at the Fermi energy $E_F$ of the electrodes. This linear relation is valid when the applied voltage bias is small.\cite{Datta95}  
It is sufficient for our applications since very low voltage (10-25 mV) is used in STM break junction experiments.\cite{tao-science,Venkataraman0600}.
Therefore the conductance for BDA and BDT junctions can be extracted from the transmission spectra plotted on Fig. \ref{fig:trans}.
The conductance of the BDA junction is 0.038 $G_0$. The experimentally measured conductance for BDA junction 
is 0.0064 $G_0$ (reported in Ref.\cite{Venkataraman0600}) that is 6 times smaller than the computed value. The reason for this discrepancy is not clear. The BDT junction shows a wide range of conductance values: 0.1 $G_0$ for geometry (h), 0.2 $G_0$ for geometry (i), and 0.7 $G_0$ for geometry (j).
Such variations are qualitatively consistent with the experimental observation that the mean conductance variation for amine anchored junction is much smaller than that for thiolate linked junction in different experimental conductance traces.\cite{Venkataraman0600}

Our study demonstrates that  BDA adsorbs on  apex Au atom, while BDT can be adsorbed on both clean Au surface and apex Au atom. It means that there are more geometrical constraints for  BDA than for  BDT in forming junctions. In addition, the N  lone-pair  leads to the stronger orientation preference  for the amine-Au bond than that for
the thiolate-Au linkage. Therefore, it is sterically more difficult to get multiple BDA molecules in a junction. 
Based on this picture, we suggest that the recently studied  dithiocarbamate and dithiocarboxylate anchoring groups, \cite{Tivanski0598,Li0616,Li0693} which are even more constrained orientationally,  would provide stable and well-defined junctions for experimental measurements.

\subsection{Conductance Trace Calculations}
We model STM break-junction experiments by our  program
for "on-the-fly" conductance calculations.
We use the most stable non-lying adsorption geometries  of phenylamine and phenylthiolate 
((d) and (j) respectively on Fig.\ref{fig:geo}) to prepare the 
cluster models for the conductance trace calculations. The atomic coordinates of the 10-atom gold cluster and the molecule are extracted from 
 the optimized adsorption geometry.  
Then we perform 150 constrained geometry relaxations for each cluster-molecule system as is described in section II B.  This series of the constrained geometry relaxations yields the trajectories which describe the evolution of the junction when the molecule is pulled from the Au
surface by the  STM tip. Each trajectory uniformly spans the junction evolution from the equilibrium stretch 0 \AA $ $ to 
150$\times$0.05 \AA = 7.5 \AA.
Fig.\ref{fig:stretch}(a) shows representative snapshots of the geometry evolutions for the BDA and BDT junctions.

 The junctions undergo completely different geometry evolutions as they are stretched. 
 The pulling of BDT causes the significant distortion of the Au cluster, which represents the electrode.
 At the same time, the geometry  of  the Au cluster is not significantly affected by the stretching  of the BDA junction.  
 This can be understood by comparing amine-Au and thiolate-Au bond strengths.
The amine-Au is a weak bond. When  BDA is pulled from the Au electrode, the bond between the amine and the apex Au atom ruptures before the molecule is able to involve the other Au atoms from the cluster into its motion. The interaction between thiolate and Au is considerably stronger than Au-Au interaction, and it involves all Au atoms into the motion as 
 the BDT junction is stretched. Therefore, the electrode geometry evolution for  the BDA junction is characterized by two states (before and after the amine-Au bond rupture), whereas the BDT junction has different electrode geometries for every stretching step.

Junction conductance traces and the geometry evolution trajectories are calculated simultaneously
 (see section II B for detail). Fig. \ref{fig:stretch} shows the 
conductance traces for the both junctions.  The conductance of BDA only slightly fluctuates around several hundredth of G$_0$ at early stage of the pulling, then it drops abruptly when the amine-Au linkage ruptures. Oppositely, 
the BDT junction undergoes  almost two orders of magnitude variation in the conductance 
 before the molecule is detached from the electrode. This difference of the behavior of the conductance variations is consistent with the experimental observations.\cite{Venkataraman0600}
We also noticed that the conductance of the BDT junction is increased as the junction is stretched by the STM tip. This behavior of the the conductance trace is in agreement with Xue and Ratner observation that the the conductance of BDT may grow  with the increase in the molecule-electrode separation.\cite{xue-ratner} It is also consistent with Ke, Baranger and Yang calculations which show that the presence of an additional Au atom 
at each of the two contacts will increase the  BDT conductance.\cite{ke-baranger-yang}

%It has not escaped from our attention that BDT conductance drops abruptly on our calculations as the thiolate-Au bond is raptured, whereas the DBT experimental shows a multistep decrease.  It could indicate that the experimental measurements of the BDT junctions are performed on array of
%With multiple molecules, experimental BDT conductance variation should be larger than simulated result by one molecule, and the conductance should be getting smaller with elongation in experiment when the number of molecule in junction decrease. We notice that in experimental conductance traces of BDT junctions, before a abrupt decrease of conductance, there may be a increase in conductance. If we consider the experimental abrupt conductance drop is caused by one BDT molecule lost it connect to electrode, then our results agree with experiment well. In fact, on the other hand, drag one more gold atom out from surface may not cause conductance decrease, as indicted by calculation for alkanedithiolate junctions with different number of apex Au. \cite{Muller0603}

\section{Conclusion}
We performed a computational study to elucidate the nature of the well-defined conductance 
of amine terminated molecular junctions. We analyzed possible adsorption geometries  of
amines and thiolates on clean and undercoordinated
Au(111) surfaces. Electron transport properties of BDT and BDA molecules are computed for 
static junctions as well as "on-the-fly" to model STM break junction  experiments.
The main results of our study can be summarized as follows:

\begin{enumerate}
\item
Amines adsorb only on undercoordinated Au surface whereas thiolates adsorb equally well on clean and undercoordinated Au surfaces. AE of the vertical phenylamine  on undercoordinated Au(111) surface is 0.50 eV larger than  AE of the most stable non-lying adsorption geometry on  clean Au(111) surface. 
AEs of phenylthiolate on on  clean and undercoordinated Au(111) surface are very similar ($\sim$ 2.7 eV).

\item
The number of accessible geometries for BDA junction  on Au surface is much smaller than that for the BDT. 
Only one geometry is sterically/energetically accessible for BDA junctions whereas three different geometries are possible 
for  BDT. It suggests it is more likely for  the amine terminated molecules to form
  single molecule junctions  than for thiolate terminated molecules.

\item
Conductance of static BDA junction is 0.0064 G$_0$ whereas conductance of BDT varies from 0.1 to 0.7 G$_0$
depending on the adsorption geometry.
\item
The conductance of BDA only slightly fluctuates at the early stage of the junction formation, then it drops abruptly as the amine-Au linkage ruptures. Oppositely,  the BDT junction undergoes  almost two order of magnitude variation in the conductance 
 before the molecule is detached from the electrode.  This is because the electrode geometry is significantly distorted by the thiolate anchored junction stretch and is not affected by the stretch of the amine anchored junction.
 
\end{enumerate}

\vspace{1cm}
\textbf{Acknowledgment.} 
The authors are grateful to M. Gelin for helpful discussion.
This work was partially supported by the American Chemical Society Petroleum Research Fund (44481-G6).


\begin{thebibliography}{99}
\bibitem{ae-bdt} 
We assume that all H atoms released upon the adsorption of phenylthiols form H$_2$ molecules. Therefore,
the chemical potential of the H atom in H$_2$ is added to the total energy of the adsorbed system, when AE of phenylthiolate is computed.
\end{thebibliography}
\end{document}